# Detection of Single Stuck-at and Bridging Faults in Cluster-based FPGA Architectures

Nachiketa Das*, Sudip Ghosh**, and H. Rahaman**

*Marine Engineering and Research Institute, Kolkata
**School of VLSI Technology, Bengal Engg. & Science University, Shibpur, Howrah – 711 103, India
Email:* nachiketad@gmail.com,** sudipghosh2005@gmail.com, **rahaman_h@yahoo.co.in

***Abstract***

*This paper presents a new technique for detecting single stuck-at and bridging faults in cluster-based FPGA architectures, using same set of test vectors generated by a test pattern generator circuitry. This technique is based on configuration of logic elements using test logic structure and the outputs of two separated logic elements are compared to detect a fault. To do this, block under test, TPG and output response analysis circuitry are generated in FPGAs. The configurations of test pattern generation, block under test and output response analysis are also discussed in this paper. The novelty lies in the fact that the same set of test vectors can detect all the stuck-at as well as bridging faults if the test vectors are applied in a particular sequence and analyzing the output results. We presented a detail analysis of results to detect faults as well.*

## 1. Introduction

Nowadays, FPGAs are widely used in many applications such as industrial, spacecraft and embedded applications. They are increasingly used in applications where previously were exclusive territory of ASICs. Today's FPGA looks like as a system on a chip [24]. It has Block RAMs, plenty of configurable logic and software design tools. FPGA is no longer a prototyping business. Electronic designers use FPGA because of its high flexibility in achieving multiple requirements such as high performance, no Non-Recurring-Engineering cost and fast Time-To-Market [19], [20]. It is notable to mention that designers, who preferred ASIC, are now using the mixed FPGA-ASIC solution [26]. FPGA reprogrammability introduces three major interesting aspects: 1) unlike ASIC designs, FPGA-based implementation does not need several manufacturing process; so, the design cycle time is reduced, 2) FPGA utilization decreases the overall design cost as compared to ASIC designs that use several expensive mask manufacturing, and 3) SRAM-based FPGAs are best suited for remote missions because of their ability to being reprogrammed by users as many times as necessary in a very short time.

Cluster-based FPGA architecture can be used to model large design on FPGA. For correctness of the stuck-at and bridging faults we need to test these FPGA architectures so that the design prototypes are totally correct and thus we can detect the above-mentioned faults in the FPGAs. Built in self test (BIST) is one of the good methods to test circuits. BIST techniques can be classified into two categories [1], [12]: 1) on-line BIST, which includes concurrent and non-concurrent techniques, and 2) off-line BIST, which includes functional and structural approaches.

We propose a method that uses BIST based testing by configuring the block under test in such a way that can detect and locate the presence of single stuck-at and bridging faults at the FPGA logic block inputs. The proposed technique will detect the presence of single stuck-at and bridging faults by the application of common test pattern. The proposed method is simulated for detection of stuck-at fault by using HOPE fault simulator [18]. We have used stuck-at fault information to detect the presence of bridging on the input of the logic block of FPGA [2], with the help of common test vector.

The rest of the paper is organized as follows. Section 2 describes related work in the field of fault detection in FPGA and various digital circuits. Section 3 will briefly describe the cluster-based FPGA architecture and the BIST based testing method. The faults that the proposed method can detect are presented in section 4. Section 5 provides the algorithm for detection of stuck-at and bridging faults. The experimental study is presented in section 6. Finally, the conclusion of this study is given in section 7.

## 2. Related work

The FPGA test problem has been divided by several researchers into the interconnect test problem [15], and the FPGA logic test problem. The limited number of I/O pads greatly reduces test access from off-chip. The pad limit has been overcome by several researchers by using a number of BIST techniques to reduce the need for pads. Some portion of FPGA hardware is configured as test pattern generation (TPG) and output response analysis circuitry (ORA), which is used to test the remainder of the FPGA. In order to test the complete FPGA, several shifting BISTER configurations are required to ensure that the entire FPGA logic is tested. Several approaches to on-line fault detection have been introduced which implement BIST by exploiting unutilized FPGA logic and routing to implement modular redundancy. The need for external controllability and observability has also been reduced by using iterative logic array (ILA) test architecture [22]. An ILA architecture composed of an array of identical cells allows the controllability and observability of each cell to be effectively accomplished through its neighboring cells.

A novel technique for using information from the stuck-at fault dictionaries to diagnose bridging faults is presented in [2]. The proposed method consists of the observation that if a test vector detects a bridging fault then it can also detect at least one stuck-at fault on the bridging node. For the test vector in the presence of the bridging fault, the observed failure data will match the failed data of the stuck–at fault. This observation of failing test vector is used to generate bridging fault candidate list, which is further reduced, by reduction of non-failing test vectors.

A differential $I_{DDQ}$ testing technique to detect and locate the FPGA bridging fault is presented by [9]. In this method no test vector for detecting fault are used. The FPGA was programmed with test configuration which sets the logic values of each interconnect. By interactively partitioning and eliminating interconnects, automated fault location was archived. Another on line testing of FPGA is presented in [12]. They have proposed a non-instructive CLB test method, which can detect any permanent structural fault in the CLB emerging at any stage during the system lifetime. It is able to correct any transient fault affecting the CLB functionality. Since it is completely implemented based on the IEEE 1149.1 infrastructure [16], no FPGA I/O pins are occupied.

BIST-based delay path testing method is described in [14]. In this work only those path that are critical for a particular application are tested. Static timing analysis is used to identify critical path, and the paths are scheduled into test sessions each of which test a set of paths in parallel.

3. Preliminaries

Cluster-based architecture [3] is consisting of array of several Basic Logic Elements (BLE) grouped to form a cluster. Fig. 1 shows the internal structure of a BLE. Each cluster consists of *N* number of BLEs with *I* number of inputs. Each cluster are surrounded by interconnects [8], [22]. The interconnect structures are a set of wire segment which can be connected by a set of switches referred to as programmable interconnect points (PIP). The cluster I/O is the input output of the cluster.

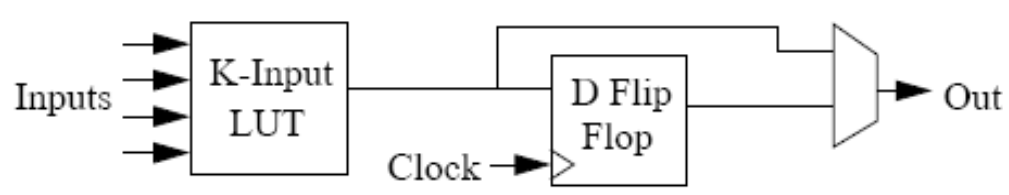

Fig. 1. Basic logic element (BLE)

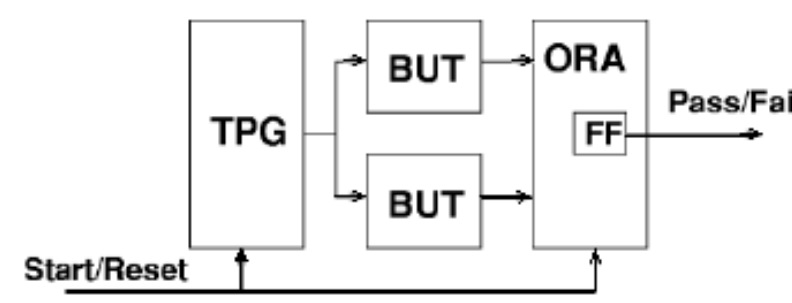

Fig. 2. BISTER test structure

Each cluster-based FPGA [3] consists of several BLEs. Each BLE consists of a Lookup table (LUT), a D Flip-flop and a Multiplexer. We can connect the output of BLE to the input of another BLE and so on to complete the cluster. All the output of BLEs collectively from the cluster outputs.

We are proposing BIST based [1], [13] testing method to test the presence of single stuck-at and bridging faults at the inputs of the LUT of a cluster. In BIST based testing method, the hardware overhead for configuring BIST over head is not the important factor, since BIST in FPGA can be easily reconfigured. To implement BIST we configure the entire FPGA with as many BISTER structures [15], [23] possible, as shown in Fig. 2. Each BISTER consists of test pattern generator (TPG), block under test (BUT) and an output response analyzer (ORA) (See Fig. 2). ORA will compare the output of two BUT in response to the test pattern generator and then it will determine the result. The TPG, ORA and BUT will be discussed latter in this paper. Many BISTER structures have to be implemented in order to cover entire FPGA [13], [15]. We can also merge the TPG and ORA together in order to increase the number of testable cells.

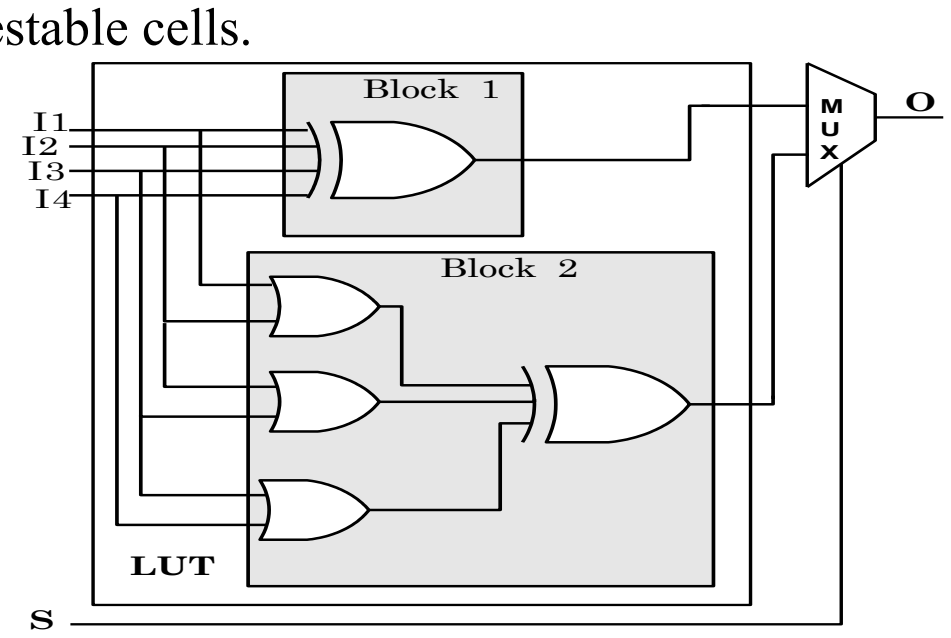

Fig. 3. Block diagram of BUT

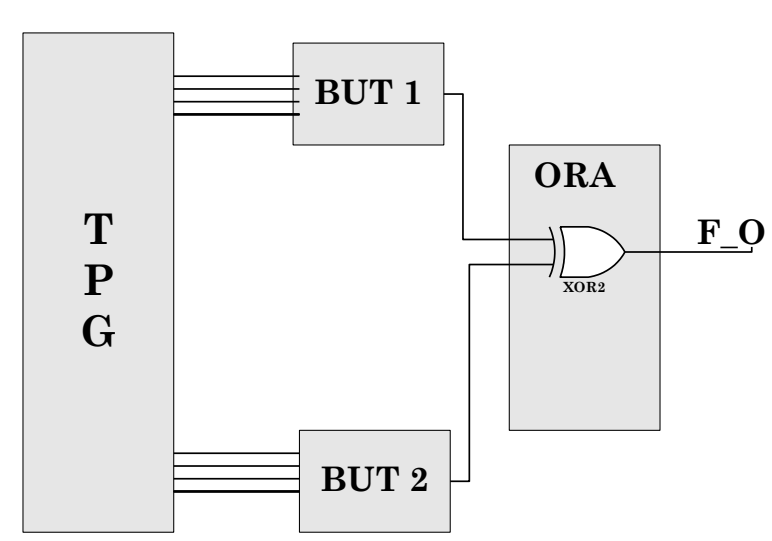

Fig. 4. Block diagram of BISTER showing ORA circuit

The clusters, which are dedicated for TPG or ORA logic, cannot be tested since it is used to test other blocks and not itself. In order to test the entire cluster, the FPGA will have to be reconfigured with the shifted BISTER structure. After repeating the shifted BISTER structure, all the LUT of a cluster can be tested except those, which are present on the perimeter of the FPGA. To test the LUT of a cluster present on the perimeter of the FPGA, the layout of the BISTER structure should be reoriented. The BUT in paper [15] was configured as XOR gate, which can only detect stuck-at fault but if we apply the test pattern as described in section 4,the BUT circuitry can also detect the presence of bridging fault. The scheme that we are proposing can detect and also locate the presence of single stuck-at and bridging faults, by the application of test pattern.

## 4. Detectable faults by the proposed scheme

To detect single stuck-at and bridging fault in a cluster-based FPGA architecture, a single BISTER structure with a common test sequence is used. These faults can be broadly classified as follows: stuck-at-0 (SA0), stuck-at-1 (SA1), AND bridged (AB), and OR bridged (OB) faults.

### 4.1. Intercluster test configuration

In order to detect the interconnection fault of the FPGA we configure the BUT, TPG and ORA blocks in such a way, that it can detect all the faults that we have discussed by a single configuration.

#### 4.1.1. BUT configuration.

We configure BLE/cluster (which will act as BUT), which we are intended to test as shown in Fig. 3. Each BLE/cluster will be configured as combinational circuit as shown in the Fig. 3. The circuit is divided into two blocks: *Block1* and *Block2*. The output of the two blocks are selected by a $2 \times 1$ multiplexer by controlling the select line *S*. *Block1* is an *I*-input XOR which can detect the present of single fault (SA0, SA1, AB, OB). If there is a presence of a single SA0, SA1 fault it will produce a faulty output in response to a particular sequence of test vectors. We can also locate the faulty node by applying the test vectors in proper sequence, which is described in the next section. If there is any presence of a bridging fault i.e., AB, OR between two nodes it will produce a faulty output. To locate the faulty nodes we have to apply test vectors in a particular order as described in the next section. *Block1* cannot tell whether the bridging fault is AB or OB, because test vectors are same for testing AB and OB and it also gives same output if any of the bridging fault is present. But *Block1* can distinguish the SA0 and SA1, because the test patterns have one unique test vector each for SA0 and SA1.

To detect the bridging fault we have to detect three separate conditions namely, the fault free condition, the presence of AB and the presence of OB. The presence of fault free and the faulty condition is detected by the *Block1*. In order to detect the type of fault, we have to give same sequence of test vector to the combinational circuit of *Block2*. It will give different output for the presence of different type of single bridging faults. Thus by comparing the result of *Block1* and *Block2* we can tell the presence of a single SA0, SA1, AB, and OB.

### 4.1.2. ORA configuration.

ORA will be configured to take input from two different BUT and compare their outputs. The ORA circuit in the figure consists of a two input XOR gate, which will accept output of two BUT, one of which is considered to be fault free. The ORA in a BISTER structure is shown in Fig.4.

### 4.2. Test pattern generation

The TPG will generate a single test pattern consisting of a combination of several test sequence. The Boolean difference method [17] is used to determine all the test sets required for testing stuck-at faults. From these test sets only those test sets are taken which can detect both AND, OR bridging faults and stuck-at faults. To differentiate between the faults whether it is stuck-at or bridging fault we will select two test sequence which will uniquely determine single stuck-at faults only. Let a function $\Gamma[x] = \Gamma[x_1, x_2, x_3, \ldots, x_i, \ldots, x_n]$ then boolean difference with respect to $x_i$ is

$$\frac{d(\Gamma[x])}{d\,\mathrm{xi}} = \Gamma\mathrm{i}[0] \oplus \Gamma\mathrm{i}[1] \qquad (1)$$

where $\Gamma_i[0]$ and $\Gamma_i[1]$ are given by

$$\Gamma_i[0] = \Gamma[x_1, x_2, x_3, \ldots, 0, \ldots, x_n] \qquad (2)$$
$$\Gamma_i[1] = \Gamma[x_1, x_2, x_3, \ldots, 1, \ldots, x_n] \qquad (3)$$

In order to test SA0 fault in $x_i$ node, we have to assign value of 1 to variable of $x_i$. This means that

$$x_i \cdot \frac{d(\Gamma[x])}{d\,\mathrm{xi}} = 1 \qquad (4)$$

This will give all test vectors which will detect SA0 fault in node $x_i$. In order to test SA1, we have to assign value of 0 to variable of $x_i$. This means that

$$\bar{x}_i \cdot \frac{d(\Gamma[x])}{d\,\mathrm{xi}} = 1 \qquad (5)$$

will give the entire test vector which will detect SA1 fault in node $x_i$. Again in case of bridging fault the two shorted lines will attain the same value. If the two lines have same value in the fault free circuit then it will not detect bridging faults. So to detect a bridging fault between two nodes, the test vector should have complimentary values in those two nodes. The test vector that can detect stuck-at fault can also detect bridging fault if the two nodes have complimentary values [2].

### 4.3. Test vector selection

From the above discussion of generating test vector, and from Fig. 5, let consider a four input LUT, we have two output functions $\Gamma_1[x]$ and $\Gamma_2[x]$ from *Block1* and *Block2* respectively which will pass through multiplexer.

$$\Gamma_1[x] = x_1 \oplus x_2 \oplus x_3 \ldots \oplus x_n \qquad (6)$$
$$\Gamma_2[x] = (x_1 + x_2) \oplus (x_2 + x_3) \ldots \oplus (x_{n-1} + x_n) \qquad (7)$$

Since *Block1* alone can detect stuck-at faults and bridging faults, *Block2* is used to distinguish between the types of bridging faults. So we will find the test sequences for *Block1* that can detect bridging fault and from those stuck-at fault sequences we will select only those sequences that can detect stuck-at faults [2].

Therefore from the Eq. (5) and Eq. (6) the test vectors that can detect SA1 fault for $x_1$ node are {(0011), (0101), (0000), (0010), (0100), (0110), (0111), (0001)}. As test vector {(0000), (0111)} can detect the presence of single stuck-at-1 fault in node $x_1$ and (0111) has complimentary value in the node $x_1$ and $x_2$ , so this test vector can detect presence of AND, OR bridge between node $x_1$ and $x_2$. Similarly all the test vectors have to be found using the method describe above. All the set of test vectors that can detect and locate SA1 faults and bridging fault in all the nodes are {(0000), (1011), (1101), (1110), (0111)}.

Similarly we will find all the test vectors that can detect and locate the presence of single SA0 fault and bridging fault by using Eq. (5) and Eq. (7) are {(1111), (1000), (0100), (0010), (0001)}.

From the test vectors we got, the test vectors {(1111), (0000)} can only detect presence of SA0 and SA1 respectively. The rest of the sequence can detect any of the presence of AND, OR bridging fault and can also locate the location of the fault. The test vector that can locate the presence of SA1 fault in $x_i$, will not response to the presence of SA0 fault in the $x_i$ node. So if the test vector (1111) detects presence of SA0 fault and if the test vector (0111) does not response to the fault then we can say the faulty node is node $x_1$. Similarly for all nodes the logic will hold. Now the total number of test vectors that can detect all the mentioned faults will be reduced to {(1111), (0000), (1011), (1101), (1110), (0111)}. In the same way the test vectors required for *I* number of input nodes can be found out.

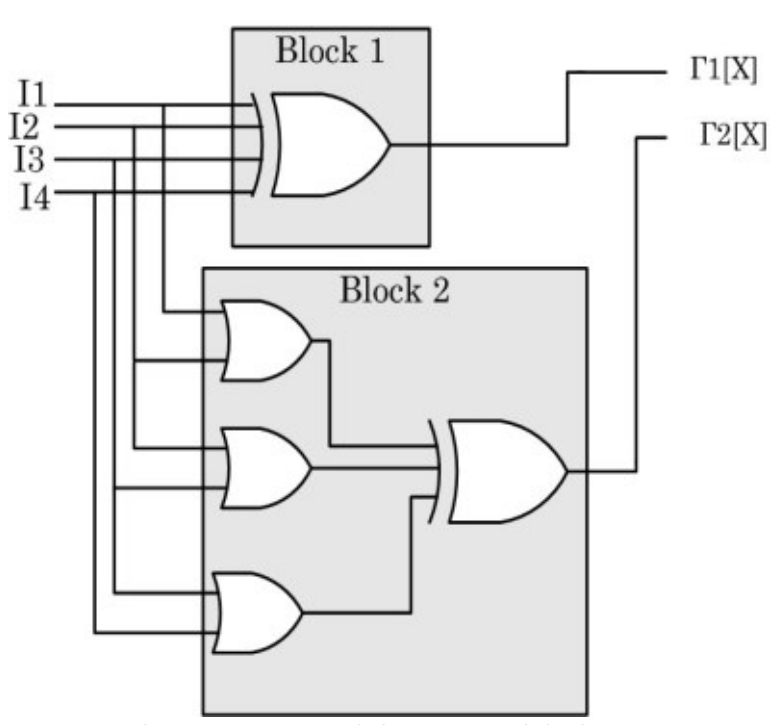

Fig 5. BUT without multiplexer

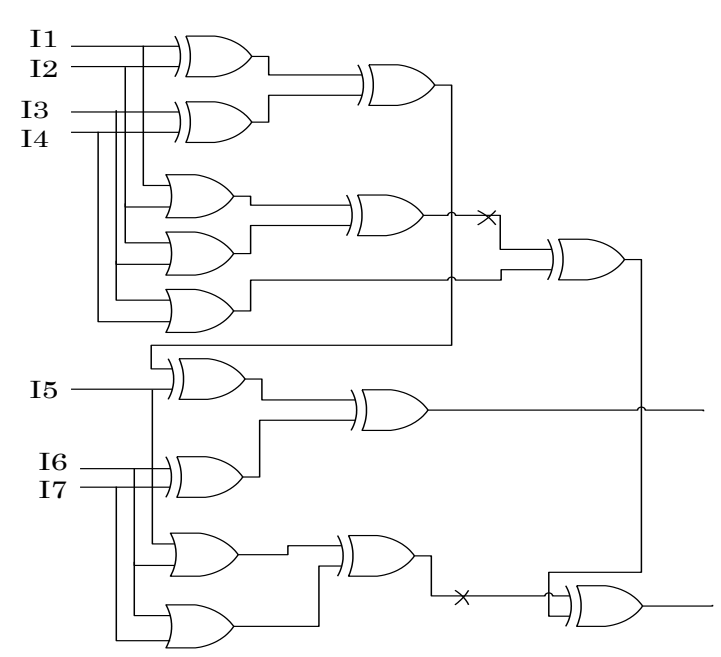

Fig. 6. Block diagram of BUT (BLOCK UNDER TEST) while simulating with HOPE (more than 4 inputs)

## 5. Detection of stuck-at and bridging faults

For simulation and finding the Percentage fault-coverage of the stuck-at faults and bridging (AND/OR) faults of the BUT of cluster-based FPGA using 'HOPE' (a fault simulator) [18], [2], we have removed the multiplexer from the circuit.

As shown in Fig. 3, the multiplexer is used for just making a selection between the two blocks of the circuit in order to test for the stuck-at fault and the bridging faults. Therefore, for simulation of the circuit using HOPE we can remove the multiplexer, then the circuit will have two outputs. Hence, for multiple LUT the circuit will be as Fig. 5 and Fig. 6. Moreover, The Slices [25] within one configurable logic block (CLB) of a cluster-based FPGA will be tested first for interconnection fault i.e., Intracluster fault detection. Then the interconnection fault between two or more CLBs of the cluster-based FPGA architecture will be tested i.e., Extracluster fault detection.

The number of inputs to a LUT (4-input LUT) of a BLE within slice of CLB will be reduced by one if we test for more than 4 input LUT, which will be formed by combining one or more 4 input LUT together i.e., if we combine $N$ number of LUT with $I$ number of input each, then we can only test $[(I-1) \times N+1]$ inputs of LUT. We have configured LUT as shown in Fig. 6.

Here *Block1* is configured as *XOR* and *Block2* is configured as a combinational logic. Let $X$ be zero value such that $X= I_1 I_2 I_3 I_4 I_5 \ldots I_n$. Here $n = 4$ is given by algorithm 1

| Algorithm 1 |
|---|
| ***if*** $S == 0$ ***then*** |
| $\Gamma[x] = OUT = [I_1 \oplus I_2 \oplus I_3 \oplus I_4]$ |
| ***else if*** $S == 1$ ***then*** |
| $\Gamma[x] = OUT = [(I_1+I_2) \oplus (I_2+I_3) \oplus (I_3+I_4)]$ |

| Algorithm 2 |
|---|
| ***if*** $S == 0$ ***then*** |
| $\Gamma[x] = OUT = [I_1 \oplus I_2 \oplus I_3 \oplus I_4] \oplus [I_5 \oplus I_6 \oplus \ldots .I_k]$ |
| ***else if*** $S == 1$ ***then*** |
| $\Gamma[x] = OUT = [(I_1+I_2) \oplus (I_2+I_3) \oplus (I_3+I_4)] \oplus [(I_5+I_6) \oplus (I_7+I_8)] \oplus \ldots \oplus (I_{n-1}+I_n)]$ |

Therefore, for $n > 4$ say $k$ inputs i.e., the BUT circuit implementation is as shown below and Fig. 7. In the case of $n = k$, it will follows algorithm 2

From the above figure only $I_6$, $I_7$, $I_8$ are testable input and $I_5$ is connected to output of previous stages. Hence it is not controllable. So from the above discussion we can say that one will reduce number of controllable and observable input if we used two LUT. If we use three LUT it will reduce by two inputs that cannot be controlled or observed. Hence for $N$ LUTs of 4 input with total inputs $I$ such that $I > 4$, number of controllable and observable inputs will be: $[(I-1) \times N+1]$.

Procedure. The ORA has two inputs which are the outputs of two BUTs. ORA compare the output of two BUTs, one of the BUT is considered being faulty and another is considered to be fault free. The equation is shown below.

$$\Gamma_O [(BUT1, BUT2)] = FINAL_OUT = BUT1 \oplus BUT2 \quad (8)$$

We have developed a set of rules to determine the bridging Faults. The rules are as follows:

### 5.1. Detection of bridging faults

To test for AND bridge we assumed that the test pattern will be of following form: The targeted input node which is to be tested is made low (say $I_1$) and other inputs are kept high, that is to test bridging fault between two node, the two node should have complimentary inputs. As we see from the Fig.4, output from the two BUT are compared for fault, as we assumed that the reference BUT should be fault free.

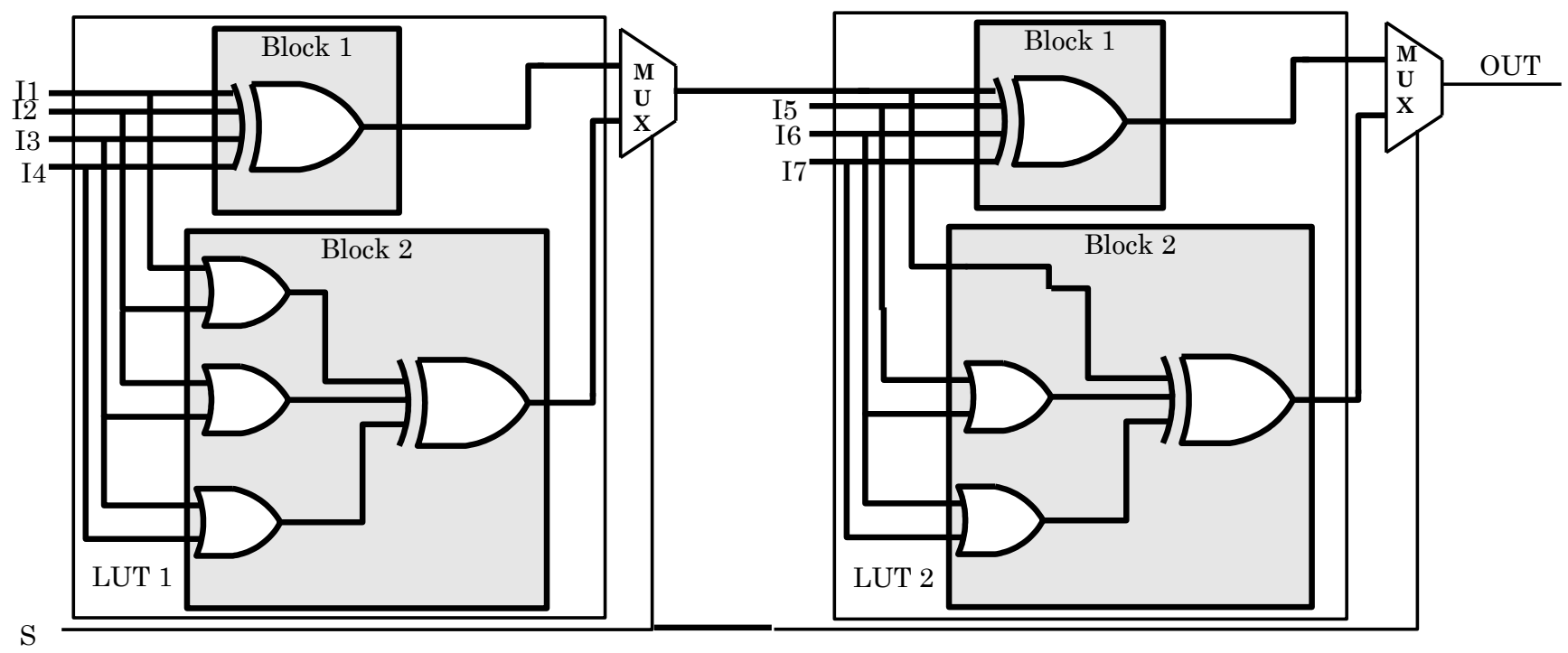

Fig. 7. Block diagram of BUT with more than 4 inputs to test

5.1.1 Detection of AND bridging fault: Assume the number of input when *I* is even. From Fig.3 and Algorithm 2 we can say

$$\Gamma[x] = OUT = 1 \qquad \text{for } S= 0 \ \& \ 1 \qquad (9)$$

From Fig. 3 and Eq. (1) can give

$$\Gamma_O[(BUT1, BUT2)] = 0 \qquad \text{for } S= 0, 1 \qquad (10)$$

However, if an AND-bridging fault is occurred in one of the BUTs, in the same case the results would be different. From Fig. 1 and Algorithm 2 we can say

$$\Gamma[x]= OUT= 0 \qquad \text{for } S= 0 \ \& \ 1 \qquad (11)$$

Similarly from Fig.2 and Eq. (1) can drive

$$\Gamma_O[(BUT1, BUT2)]= FINAL_OUT= 1 \quad \text{for } S= 0, 1 \qquad (12)$$

5.1.1 Detection of OR bridging fault: Again, if there is no bridging fault, these following results will be correct. The number of input when *I* is even. From Fig.3 and Algorithm 2 we can say

$$\Gamma[x]= OUT= 0 \qquad \text{for } S= 0, 1 \qquad (13)$$

From Fig.2 and equation 1

$$\Gamma_O[(BUT1, BUT2)]= 0 \qquad \text{for } S= 0, 1 \qquad (14)$$

However, if an OR-bridging fault is occurred in one of the BUTs, in the same case the results would be different. The number of input when *I* is odd. From Fig.7 and Algorithm 2 we can say

$$\Gamma[x]= OUT= 1 \qquad \text{for } S= 0,1 \qquad (15)$$

Similarly from Fig.2 and Eq. (1)

$\Gamma_O[(BUT1, BUT2)]=1 \quad for\ S= 0\ \&\ 1$ (16)

Hence from the above assumption, number of test vectors required to test *I* inputs will be *I* of *[I+1]* bit length. Again as we assume that only one fault exist at a time between two consecutive input pins, hence the number of test vectors will be now *[I+2]*.

5.2. Detection of stuck-at-1 fault

Assumption: To test stuck-at-1 first input test vector will be all low and to detect the faulty pin we make the input of the targeted node below and the rest of the node all high. Let $X_1= I_1\ I_2\ I_3\ I_4\ I_5\ ....I_n$ and $X_2= I_1\ I_2\ I_3\ I_4\ I_5\ ....I_{n.}$ Say $I_1$ is the target node then from Algorithm 1 and Fig.3. Suppose the no stuck-at-1 error is in the BUT. Hence, in the case of *S* equals to 0, following equations are exists

$\Gamma\ [X_1] = OUT_{X1} = 0 \quad for\ X_1$ (17)

$\Gamma\ [X_2] = OUT_{X2} = 0 \quad for\ X_2, I\ is\ odd$ (18)

$= 1 \quad for\ X_2, I\ is\ even$

$\Gamma_O\ [(BUT1_{X1}, BUT2_{X1})] = 0$ (19)

$\Gamma_O\ [(BUT1_{X2}, BUT2_{X2})] = 0$ (20)

However, if an stuck-at-1 error happened in a BUT, the following wrong results will be generated

$\Gamma\ [X_1] = OUT_{X1} = 1 \quad for\ X_1$ (21)

$\Gamma\ [X_2] = OUT_{X2} = 1 \quad for\ X_2, I\ is\ odd$ (22)

$= 0 \quad for\ X_2, I\ is\ even$

$\Gamma_O\ [(BUT1_{X1}, BUT2_{X1})] = 1$ (23)

$\Gamma_O\ [(BUT1_{X2}, BUT2_{X2})] = 1$ (24)

So based on the wrong results generated by the stuck-at-1 error, we can mention a method to detect stuck-at-1 error as show in algorithm 4.

*Algorithm 3 /*Bridging fault detection*/*
*In the case of S=0 {*
*Result =* $\Gamma_O$ *[(BUT1, BUT2)]*
*If Result1 is False then*
*AND/OR bridging fault detected*
*else if Result is False then*
*No AND/OR bridging detected*
*}*
*In the case of S=1 {*
*Result=*$\Gamma_O$*[(BUT1, BUT2)]*
*If Result1 is False and Result2 is False then*
*AND bridging fault detected*
*Else if Result1 is False and Result2 is True then*
*OR bridging fault detected*
*}*

Algorithm 4 */*Stuck-at-1 fault detection*/*
*In the case of S=0 {*
*Result1=* $\Gamma_O$ *[(BUT1* $_{X1}$*, BUT2* $_{X1}$*)]*
*Result2=* $\Gamma_O$ *[(BUT1*$_{X2}$*, BUT2* $_{X2}$*)]*
*If Result1 is True and Result2 is True then*
*No Stuck-at-1 fault occurred*
*Else if Result1 is False and Result2 is False then*
*Stuck-at-1 fault on the faulty pin is I of* $X_2$ *occurred*
*}*

### 5.3. Detection of stuck-at-0 fault

Assumption: To test stuck-at-1, first input test vector will be all low and to locate the faulty node we make the input of the targeted pin low and the rest of the pins all high. Let $X_1$= ($I_1$, $I_2$, …, $I_n$), $X_2$= ($I_1$, $I_2$, …, $I_n$) and $X_3$= ($I_1$, $I_2$, …, $I_n$).

Before stuck-at-0 happened in the BUT, in the case of *S* is zero, following results are generated by the circuit shown in Fig. 7.

$\Gamma\,[X_1] = OUT_{X1} = 0$ for $X_1$ (25)
$\Gamma\,[X_2] = OUT_{X2} = 0$ for $X_2$, $I$ is odd (26)
$= 1$ for $X_2$, $I$ is even
$\Gamma\,[X_3] = OUT_{X3} = 0$ for $X_2$, $I$ is odd (27)
$= 1$ for $X_2$, $I$ is even
$\Gamma_O\,[(BUT1_{X1}, BUT2_{X1})] = 0$ (28)
$\Gamma_O\,[(BUT1_{X2}, BUT2_{X2})] = 0$ (29)
$\Gamma_O\,[(BUT1_{X3}, BUT2_{X3})] = 0$ (30)

However, suppose that a given stuck-at-0 happened in the BUT, so some wrong results may be generated in the case of *S* is zero*,* as show in follows:

$\Gamma\,[X_1] = OUT_{X1} = 1$ for $X_1$ (31)
$\Gamma\,[X_2] = OUT_{X2} = 0$ for $X_2$, $I$ is odd (32)
$= 1$ for $X_2$, $I$ is even
$\Gamma\,[X_3] = OUT_{X3} = 1$ for $X_2$, I is odd (33)
$= 0$ for $X_2$, I is even
$\Gamma_O\,[(BUT1_{X1}, BUT2_{X1})] = 1$ (34)

$\Gamma_O\,[(BUT1_{X2}, BUT2_{\,X2})] = 1$ (35)
$\Gamma_O\,[(BUT1_{X3}, BUT2_{\,X3})] = 1$ (36)

Therefore, based on these mentioned equations, when a stuck-at-zero is happened in the BUT, it can be detected by comparing correct results and wrong results as show in the algorithm 5. Therefore, total number of test vector required to test the AND, OR bridging and stuck-at fault shall be (*I*+2) where *I* is the number of inputs of LUT that can be tested.

```
Algorithm 5        /*Stuck-at-0 fault detection*/
 In the case if S = 0 {
      Result1 = Γ_O [(BUT1 _X1, BUT2 _X1)]
      Result2= Γ_O [(BUT1_X2, BUT2 _X2)]
      Result3= Γ_O [(BUT1_X3, BUT2 _X3)]
      If Result1 is  True and Result2 is  True
          and Result3 is True then
             No  Stuck-at-0 fault occurred
      Else if Result1 is False and Result2 is False
             and Result3 is True then
             Stuck-at-0 fault an the faulty pin is I of X_3 occurred
}
```

6. Experimental results

The proposed circuit is converted in ISCAS89 netlist format. The method described in [2] is used to detect bridging faults using HOPE fault simulation tools [18] and MATLAB Tool. The circuit is applied in *.bench* format, the faults to be tested are applied in *.fault* format and the test vectors are applied in *.test* format. The fault table and the observed table are made as described in the [2]. The algorithm is described in the [2] is implemented using MATLAB. The fault table and observed table are applied as input to the program to know whether the applied test vector detects bridging faults. For detection of stuck-at fault the circuit is applied in *.bench* format along with *.fault* and *.test*. Entire experiment is repeated for various input and the LUT. The results are tabulated in table 1.

The first column denotes the number of input to the circuit. The second column denotes the number of test vectors required. The third column is divided into two parts, one part denotes percentage fault coverage for stuck-at fault and another part for the fault coverage of bridging faults. The fourth column denotes the number of Flip-Flops of BUT. The fault detection coverage is like [13] with 100% however, we can also detect bridging faults with coverage of 100% while work [13] is not proposed for bridging faults.

Faults that are detectable by this method: we are successfully detecting the entire mentioned faults at the testable inputs of the LUT (cluster).

Faults that are not detectable by this method: form the fig-6 this method cannot detect stuck-at 0 fault at the line having cross mark. Since in this portion of the circuits fit inside the LUT of a cluster /slice so it will not affect the total fault coverage. The total number undetected line twenty five inputs are eight, and for seven inputs are two. Hence we are getting total percentage of fault coverage as 100%.

**Table1. The experimented results**

| No. of Inputs of BUT | Test Pattern | Fault Coverage | | No. of FFs in BUT |
|---|---|---|---|---|
| | | Bridging | Stuck-at | |
| **4** | **6** | **100%** | **100%** | **0** |
| **7** | **9** | **100%** | **100%** | **0** |
| **13** | **15** | **100%** | **100%** | **0** |
| **25** | **27** | **100%** | **100%** | **0** |

7. Conclusions

In this paper, we proposed a new method to detect single stuck-at and bridging faults in the BLE internal routings. These faults are categorized in four different ones stuck-at-0, stuck-at-1, OR bridge and AND bridge faults. A BUT configurations was proposed so that these mentioned faults can be detected based on comparing the results of fault-free and a faulty BLE. Based on the mentioned BLE configuration, several algorithms were presented that can detect and locate these faults inside of BLEs. The experimental results shown that this method can detect internal stuck-at and bridging fault with 100 percent of coverage.